\documentstyle[12pt]{article}
\begin{document}

\title{ Amplification and Disorder Effects on the Coherent Backscattering in a
Kronig-Penney Chain of Active Potentials }
\author{N. Zekri$^{1,2}$, H. Bahlouli$^{1,3}$ and Asok K. Sen$^{1,4}$  \\
\small $^{1}$International Center for Theoretical Physics, 34100 Trieste, Italy. \\
\small $^{2}$U.S.T.O., Departement de Physique, L.E.P.M., \\
\small B.P.1505 El M'Naouar, Oran, Algerie.  \\
\small $^{3}$Physics Department, King Fahd University of Petroleum
and Minerals, \\
\small Dhahran 31261, Saudi Arabia. \\
\small $^{4}$LTP Division, Saha Institute of Nuclear Physics, 1/AF Bidhannagar, \\
\small Calcutta 700 064, India.  }

\maketitle

\newpage

\begin{abstract}
\hspace{0.33in} We report in this paper the analytical and numerical results on \
the effect of amplification on the transmission and reflection coefficient of a
periodic one-dimensional Kronig-Penney lattice. A qualitative agreement is found
with the tight-binding model where the transmission and reflection increase for small
lengths before strongly oscillating with a maximum at a certain length. For larger
lengths the transmission decays exponentially with the same rate as in the growing
region while the reflection saturates at a high value. However, the maximum
transmission (and reflection) moves to larger lengths and diverges in the limit
of vanishing amplification instead of going to unity.  In very large samples, it
is anticipated that the presence of disorder and the associated length scale 
will limit this uninhibited growth in amplification.  Also, there are other
interesting competitive effects between disorder and localization giving rise
to some nonmonotonic behavior in the peak of transmission.

\end{abstract}
\vspace{0.2in}


\newpage

\noindent{\bf Introduction }

\hspace{0.33in}
Recently, there has been a lot of interest in non-hermitian hamiltonians and
quantum phase transitions (typically localized to extended wavefunctions) in
systems characterized by them.  There are in general two classes of problems
in this context: one in which the non-hermiticity is in the nonlocal part
\cite{nlnh1,nlnh2} and the other in which it is in the local part [3-8].
In the first category, one considers an imaginary vector potential added
to the momentum operator in the Schrodinger hamiltonian and this is shown
to represent the physics of vortex lines pinned by columnar defects where
the depinning is achieved \cite{nlnh1} by a sufficiently high transverse
magnetic field.  In the case of a tight-binding hamiltonian, the
non-hermiticity is introduced by a directed hopping in one of the
directions (or more), and again in this case, it is intuitively clear that
delocalization may be obtained in the preferred direction in the presence
of randomness in the local potential even in 1D.  In the second category
(non-hermiticity in the local term), an imaginary term is introduced in
the one-body potential.  It is well-known from textbooks on quantum mechanics
that depending on the sign of the imaginary term, this means the presence
of a sink (absorber) or a source (amplifier) in the system. It may be
noted that this second category does also have a counterpart in classical
systems characterized by a Helmholtz (scalar) wave equation as well, where
the practical application is in the studies of the effects of classical wave
(light) localization due to backscattering
in the presence of an amplifying (lasing) medium that has a complex
dielectric constant with spatial disorder in its real part \cite{pradhan,zhao}.
There is a common thread binding both the problems though, namely that the
spectrum for both becomes complex (the hamiltonian being non-hermitean or
real non-symmetric), but can admit of real eigenvalues as well.  The
common property is that the real eigenvalues represent localized states
and the eigenvalues off the real lines extended states.  That it is so
in the first category has been shown in the recent works starting with
Hatano and Nelson and followed by others \cite{nlnh1,nlnh2}.  For the
second category with sources at each scatterer and in the absence of
impurities, it seems counter-intuitive that there are localized
solutions; but it has been shown in a simple way \cite{sen} that the real
eigenvalues are always localized. At present,  there is no unified analysis
of non-hermiticity of both types present. In the rest of the paper we would
be concerned with non-hermitean hamiltonians of the second category only.

\hspace{0.33in} The interest in amplification effects of classical and quantum waves
in disordered media has been strongly motivated by the recent experimental
results on the amplification of light \cite{genack}.
The amplification was shown to strongly enhance the coherent backscattering and
consequently increases the reflection [1-3]. These results
on the reflection naturally lead
us also to predict an enhancement of the transmission in such amplifying systems
(which has not been examined in previous works). However, recently Sen \cite{sen}
found for periodic systems that the transmission coefficient starts increasing
exponentially up to a certain length scale where it reaches its maximum, then it oscillates
strongly before decaying at larger length scales.
The reflection seems to saturate to a constant value and becomes large asymptotically. In this
paper, we study in details both analytically and numerically this scaling
behavior of the reflection and transmission within the framework of the
Kronig-Penney model which differs from the tight-binding one by the fact that
it is a continuous multiband model where the bandwidth depends on the potential
strength while the tight-binding (TB) framework is a discrete single band model where
the bandwidth does not depend on the energy site. We compare the results with those
obtained by Sen \cite{sen}
in the tight-binding model and study the evolution of this behavior with
amplification. The competition effect between amplification and disorder is also examined.

\noindent{\bf The Model }

\hspace{0.33in} We consider a non interacting electron moving in a periodic
system of $\delta$-peak potentials of a complex strength
$\lambda=\lambda_{0} + i \eta$ where both $\lambda_0$ and $\eta$ are constant numbers.
By using the Poincare map \cite{Bellis},
the Schr\"{o}dinger equation of this system can be transformed to the following
discrete second order equation \cite{zekri}
\begin{equation}
\psi_{n+1} + \psi_{n-1}=\Omega \psi_{n}~, \label{discrete}
\end{equation}
where $\psi_{n}$ stands for the electron wavefunction at the site $n$ and
\begin{equation}
\Omega=2 {\rm cos}(\sqrt{E}) + \lambda \frac{{\rm sin}(\sqrt{E})}{\sqrt{E}} \\
= 2 {\rm cos}(k), \label {omega}
\end{equation}
where $k$ is the wave-number. In the passive lattice ($\lambda$ is real) the
corresponding wave-number is imaginary in the allowed band
($\left|\Omega\right| \leq 2$) and the wavefunction becomes Bloch like
while in the band gap it becomes real and the wavefunction is evanescent.
In the case of the active lattice ($\lambda$ is complex) the wave-number becomes
complex ($k = k_{s} + i\gamma$) and Eq.(\ref{omega}) yields
\begin{eqnarray}
2 {\rm cos}(\sqrt{E}) + \lambda_{0} \frac{{\rm sin}(\sqrt{E})}{\sqrt{E}}=
(e^{\gamma}+e^{-\gamma}) {\rm cos}(k_{s}), \\
\eta \frac{{\rm sin}(\sqrt{E})}{\sqrt{E}} = (e^{-\gamma}-e^{\gamma}){\rm sin}(k_{s}).
\label{compl}
\end{eqnarray}
The main difference between the tight-binding and this model is the direct dependence of the
amplifying term $\gamma$ on the electronic energy. If we restrict ourselves to the
first band ($0 < k_{s} < 2\pi$) we see from (4) that $\gamma$ is negative if $\eta$ is
positive. Obviously, in successive bands the sign of $\eta$ must be changed alternatively
to get the same sign of $\gamma$.
We note also that since we choose in our model, for initial conditions of the
discrete equation (\ref{discrete}), an electron moving from the right side to
the left side of the sample (see ref. \cite{zekri}) the amplification should
occur for negative values of $\gamma$. Therefore the imaginary part of the
potential should be positive in the first allowed band of the corresponding
passive system.
Indeed, in the passive system the Hamiltonian is time reversal invariant but not
in the active one, since the Hamiltonian is not hermitian.
>From Eq.(\ref{discrete}) the transmission coefficient can be obtained as
\begin{equation}
T=\frac{4{\rm sin}^{2}(\sqrt{E})\left|e^{ik_{s}}e^{-\gamma}-e^{ik_{s}}e^{\gamma}
\right|^{2}}
{\left|c e^{ik_{s}L} e^{-\gamma L} - d e^{-ik_{s}L}e^{\gamma L}\right|^{2}},
\end{equation}
and the reflection coefficient
\begin{equation}
R=\frac{\left|a e^{ik_{s}L}e^{-\gamma L}-b e^{-ik_{s}L}e^{\gamma L} \right|
^2}{\left|c e^{ik_{s}L}e^{-\gamma L} - d e^{-ik_{s}L} e^{\gamma L} \right|^2},
\end{equation}

where
\begin{eqnarray}
a= \left[ e^{i(k_{s}-\sqrt{E})}e^{-\gamma}-1 \right] \left[ e^{i(k_{s}+\sqrt{E})}
e^{-\gamma}-1 \right], \\
b= \left[ e^{-i(k_{s}+\sqrt{E})}e^{\gamma}-1 \right] \left[ e^{-i(k_{s}-\sqrt{E})}
e^{\gamma}-1 \right], \\
c=2 - \left[ e^{i(k_{s}-\sqrt{E})} e^{-\gamma}+e^{-i(k_{s}-\sqrt{E})} e^{\gamma}
\right], \\
d=2 -\left[ e^{i(k_{s}+\sqrt{E})} e^{-\gamma} + e^{-i(k_{s}+\sqrt{E})}
e^{\gamma} \right].
\label{coef}
\end{eqnarray}
Since we are interested to scan the growing and decaying regions of the
transmission coefficient (and also the reflection coefficient) it turns out
 to be more efficient to write the coefficients $c$ and $d$ as follows
\begin{equation}
c=e^{-ik_{s} \theta_{c}}e^{\gamma L_{0}} \,\,\,\, , \,\,\,\,
d=e^{ik_{s} \theta_{d}}e^{-\gamma L_{1}}, \label{coef2}
\end{equation}
where
\begin{equation}
L_{0}=\frac{{\rm ln}2 \left[{\rm cosh}(\gamma)-{\rm cos}(k_{s}-\sqrt{E}) \right]}{\gamma} \,\,\, , \,\,\,
L_{1}=-\frac{{\rm ln}2 \left[{\rm cosh}(\gamma)-{\rm cos}(k_{s}+\sqrt{E}) \right]}{\gamma}~, \label{lengths}
\end{equation}
and $\theta_{c,d}$ are real phase parameters, which are expected to
contribute to the oscillations of $T$, and behave linearly in
$\gamma$ for vanishing amplification. The transmission then reads
\begin{equation}
T=\frac{4{\rm sin}^{2}(\sqrt{E}) \left|e^{ik_{s}}e^{-\gamma}-e^{-ik_{s}}e^{\gamma}
 \right|^{2}}{\left| e^{ i(k_{s}L-\theta_{c})}
e^{-\gamma(L-L_{0})} - e^{ -i(k_{s}L-\theta_{d})}
e^{ \gamma(L-L_{1})} \right|^{2}}~. \label{trans}
\end{equation}
\\
\noindent{\bf Results and Discussion }

\hspace{0.33in} From Eqs. (3 and 4) the amplification rate $\gamma$ depends
explicitely on the potential strength and the energy. However, since we are interested
on the effect of $\gamma$ on the transmission and reflection, we can, without
loss of generality, fix the energy and the real part of the potential. The amplification
will then depend on the imaginary part of the potential. In the rest of
the text we take $E=1$ and $\lambda_{0}=0$ except for the disordered case where
$\lambda_{0}$ is taken to be uniformly distributed in the domain [$-W/2,W/2$]
where $W$ is considered as the disorder strength.
The decay of $T$ for an absorbing chain
is found from the above equations to be qualitatively similar to that for a
disordered chain (with $\eta=0$).
Thus, nothing particularly interesting takes
place for absorbers.  But, as we discuss below, in the amplifying chain there
is an interesting competition between amplification and disorder in the small
length scale regime.  So our study below focusses on the amplification where
$\eta$ must be positive. For the numerical calculations, it is easier to use $\eta$
instead of $\gamma$. In the limit of small $\gamma$ we have $\eta = -2 \gamma$.

\hspace{0.33in} In figure 1, we show the transmission as a function of the sample
length for two different amplifications. It is shown that the transmission grows
exponentially up to an oscillatory region where it assumes a maximum value. For
much larger lengths the transmission decays exponentially as in the case of an
absorbing chain. A similar behavior is shown in figure 2 for the reflection
coefficient where in contrast to the transmission, for large lengths the
backscattering saturates (instead of decaying) at a high value of the reflection 
coefficient. This behavior is in a close agreement with that of the
TB model \cite{sen} with a slight difference in the oscillatory
region due to the different dependence of $\gamma$ on $\eta$. This means that this
effect is globally model independent. It is also shown from these figures that the
maximum transmission and reflection increase by decreasing $\eta$ and shift
to higher sample lengths. Indeed, from Eq.(\ref{trans}) we see that when $L < L_{1}$
the coefficient $d$ becomes dominant and then $T$ behaves as
exp$(2 \left| \gamma \right| L)$ while at asymptotically large lengths, the coefficient
$c$ becomes dominant and the transmission decays as exp$(-2 \left| \gamma \right| L)$.
In the oscillatory region the two coefficients $c$ and $d$ are of the same order and
the length of maximum transmission is
\begin{equation}
L_{max}=\frac{1}{\gamma}{\rm ln}\frac{{\rm cosh}(\gamma)-{\rm cos}(k_{s}-\sqrt{E})}
{{\rm cosh}(\gamma)-{\rm cos}(k_{s}+\sqrt{E})}~. \label{lmax}
\end{equation}
It is clear from this equation that $L_{max}$ diverges for vanishing $\gamma$.
However, since the maximum transmission must be naturally unity for a passive
medium, $T_{max}$ should not diverge for $\gamma$ exactly equal to zero.
Thus there is an infinite discontinuity at $\eta=0$ which should turn towards
a finite discontinuity at a finite disorder $W>0$.  In order to examine the
limiting behavior as $\eta \to 0$, let us use a perturbative treatment for
$\eta \ll 1$. In this limit $k_{s}$ tends to $\sqrt{E}$ as
\begin{equation}
k_{s} = \sqrt{E} + \frac{\gamma ^{2}}{2{\rm tan}(1)}~,
\end{equation}
and from Eq.(\ref{lengths}) the lengths $L_{0}$ and $L_{1}$ are given by
\begin{equation}
L_{0} = \frac{{\rm ln}(\gamma^{2})}{\gamma} \,\,\,\,\, , \,\,\,\,\,
L_{1} = - \frac{{\rm ln}(4{\rm sin}^{2}k_{s})}{\gamma}~.
\end{equation}
It may be noted that for very small $\eta$, $T$ initially increases extremely
slowly with $L$ until it comes quite close to $L_1$, and then it shoots up
very fast to a very large value of peak transmission given by
\begin{equation}
T_{max} = \frac{1}{\gamma^{2}}~,
\end{equation}
and the length where this highest peak is obtained is given by
\begin{equation}
L_{max} = \frac{{\rm ln}(\gamma^{2} / {\rm sin}^{2}k_{s})}{2 \gamma}~,
\label{nlmax}
\end{equation}
with the proviso that a negative value of $L_{max}$ indicates that the peak is 
only at $L_{max}=0$.  Obviously this divergence with a discontinuity is a
somewhat unexpected behavior of the transmission. This is due to the fact that
when $\eta \to 0^+$, $L_{max}$ diverges faster than the amplification
length-scale $l_a = 1 / \gamma$. Therefore $\gamma L_{max}$ will also diverge
and whenever $\gamma$ is different from zero (positive), the current grows
slowly up to a very large length scale and reaches very high values.  One may
note that the asymptotic reflection coefficient $R(L=\infty)$ also diverges as
$\eta \to 0^+$ and has an infinite discontinuity at $\eta=0$.  Hence there is an
extremely high amplification in the backscattered wave for a very small $\eta$.
For example, for a chain with $\eta=10^{-4}$, $E=1.0$, the transmission peak
occurs at $L_{max}=2.07*10^5$, and $T_{max}=2.87*10^{10}$, and the asymptotic
$R(L=\infty)=1.13*10^9$ which occurs at $L>L_{max}$. 
It is also seen from Figs. 1 and 2 that the period of the oscillations
increases when $\gamma$ decreases due to the increase of $k_{s}$.  Before
passing on we would like to mention that all the effects discussed above
appears qualitatively similarly in the TB model as well.  For simplicity, if we
take the Fermi energy at the band-center ($E$ = 0),  then we find that the maximum
peak for transmission occurs at an $L_{max} \simeq 1/\eta~\rm{ln}(8\pi/\eta)$
which clearly diverges with $|\eta| \to 0$ and so does $T_{max}$.

\hspace{0.33in} However, the high amplitude of the largest peak in the
transmission or the asymptotic value of the reflection coefficient even for very
small amplification may not be observed experimentally since it occurs at very
large sizes (see Eq.(\ref{nlmax})) and the experimental realization of such
perfect (disorder-free) systems is very difficult.  Disorder, however small,
would be present (in such a very large size system) and this may cut down
strongly the divergences mentioned above.  Now, as soon as one introduces
disorder or, rather takes care of the disorder, however small, the question
regarding whether we should average or not comes up.  On the one hand it is
clear that experimentalists work on a typical sample, and not on a hypothetical
`average' sample.  On the other hand, it may not be easy to keep a sample in
the same state for a long time due to different types of relaxation processes.
Thus, the sample may change its characteristic with time if the characteristic
under consideration is highly configuration dependent.  Below we discuss both
the non-averaged and averaged transmission properties.

First, we discuss the properties for a particular configuration.  For this part, we
keep the disorder strength constant at $W=1$.  In Fig.3 we show the
effect of disorder on the transmission for different imaginary potentials. We see
clearly that the disorder destroys the amplification at larger scales and shifts
the maximum transmission to smaller lengths. The transmission fluctuations
appearing in Fig.3 increase with the amplification ($\eta$).  As is well known,
disorder introduces an exponential decay of the transmission with a rate
$\gamma _{dis} = W^{2}/96E$ \cite{souk} where $E$ is the energy of the incoming
electron and $\gamma _{dis}$ is the Lyapunov exponent due to the disorder.  Stated
differently, disorder introduces the localization length $\xi_{dis}=1/\gamma_{dis}$
into the problem.  For a small $\eta$, the length $L_{max}$ up to which the
exponential growth occurs in pure systems may be much larger than $\xi_{dis}$. So,
in general, the transmission starts decaying due to disorder effects before it
gets the maximal amplification due to a non-zero $\eta$.  Therefore, the
divergence in $T_{max}$ observed in periodic systems disappears with the included
disorder as shown in the Fig.4.  For very small $\eta$, $T_{max}$ tends to
the trivial constant value of unity with $L_{max}=0$.  But we have to remember
that for $\xi_{dis}<L<L_{max}$ (for pure systems), there is a fine competition
between the amplification-dependent growth and disorder-dependent decay which
affects the transmission sensitively in this regime.  As given by the above
formula, for $W=1$, $\xi_{dis} \simeq 100$.  Yet, indeed, there is a non-monotonic
behavior at much larger lengths corresponding to some compensation between
disorder and amplification.  For $\eta \simeq 10^{-3}$, the transmission in
general decays for $L > \xi_{dis}$ but only to pick up again at a still larger $L$,  
and one observes a peak of $T_{max}$ (for the particular disorder configuration
in Figs. 3 and 4)) at $L \simeq 260$.  This transmission peak seems to correspond
to one of the Azbel resonances that becomes sensitively amplified by a tuned value
of $\eta \simeq 10^{-3}$.  We have actually checked that this resonance peak
$T_{max}$ occurs at the same $L_{max}$ but becomes weaker both by increasing or
by decreasing $\eta$ around 0.001 as shown in Fig.4 and thus $T_{max}$ has a
peak close to this special value of 0.001 for this particular configuration.  In
particular, if we decrease $\eta \to 10^{-6}$, the peak remains at $L_{max}
\simeq 260$ while $T_{max} \to 1$ continuously.  For $\eta < 10^{-6}$, the (local)
peak transmission at $L \simeq 260$ becomes less than unity and hence the global
$T_{max}=1$ (trivial constant) and $L_{max}$ jumps back to the trivial value of
zero discontinuously (see the insert of Fig.4).  Further, as expected, we found
that in other configurations, the peak in $T_{max}$ at the special value of
$\eta \simeq 10^{-3}$ as shown in Fig.4 does not exist. 

Next we discuss the characteristics of averaged samples.  The question of what
quantity to average becomes crucial now.  In Fig.5, we choose $E=1$, $W=0.01$
and $\eta=0.1$ and show the transmission as a function of $L$ (in semi-log plot)
by averaging in (a) the quantity $T$ itself, and in (b) the quantity ln$T$.  For
comparison we have also shown the case without disorder by dashed lines.  The full
line is the result of averaging with 100 configurations and the dotted line is
the same for 10000 configurations in both the cases.  Whereas in Fig.5(a) the
average with 10000 configurations lies higher than that with 100 configurations
(both of them larger than the pure case as well!), the logarithmic average
shown in Fig.5(b) is much more well-behaved in every respect.  The results 
shown here are consistent with the fact that all the moments of the transmission
and reflection diverge in the amplifying case \cite{frei,beenaker}.  So, we
restrict ourselves to logarithmic averaging.  In the Fig.6, we show such
averaged $T_{max}$ for two different $\eta$'s (0.01 with open squares, and 0.1
with crosses).  To show both the cases with very different $T_{max}$'s we have
normalized both of them by their values for the pure case.  Now, one expects
that the nonmonotonic behavior as seen above should disappear since the Azbel
resonances disappear on averaging.  But, interestingly enough the fine tuning
between disorder and amplification is still at work, and some nonmonotonic
effects still survive.  We have shown in the inset of Fig.6 the magnified view
of the $y$-axis around 1.  Now we find that for the case of $\eta=0.1$,
there are some values of disorder $W$ around 0.01 where the $T_{max}$ is
somewhat larger than its value for the pure case.  Further, we could not find
such an interesting non-monotonic behavior for the case of $\eta=0.01$ after
a lot of search, which means that even if it is there it is probably very
weak or lies in an extremely narrow region for this case.  In any case,
Fig.6 shows amply that the fine tuning between disorder and amplification
may lead to quite interesting and unexpected results. 

\noindent{\bf Conclusion }

\hspace{0.33in} We have studied in this paper, within the framework of the
Kronig-Penney model, the effect of amplification on the transmission and
reflection of a periodic system. The behavior shown is in a close agreement with
that shown in the tight-binding model \cite{sen}. Therefore, this effect seems
to be model independent. However, this result means that diverging transmission
will be obtained at very large sample sizes for vanishing amplification while
in a passive system the transmission and reflection do not exceed one. This
limiting effect is due to the divergence of the maximum transmission length
faster than $1/ \gamma$. This effect is probably experimentally irrealizable
since large periodic samples cannot be growth without disorder which can destroy
this divergence. Indeed, we found that the maximum transmission decreases when
amplification decreases and tends to sature at one for vanishing $\gamma$. A peak
at $\eta = 0.001$ appears and seems to correspond to the compensation between
disorder and amplification. However, the decay of the transmission is slower for
smaller amplification leading to the delocalization of the electronic states.
\hspace{0.33in} However, the compensation effect leading to some of the non-monotonic
behavior in Fig.4 persist by averaging in Fig.6 and remain not well understood.
Therefore they should be extensively examined. Also the
generalization of this study to different electron energies and non-zero real
parts of the potential ($\lambda_{0}$) is necessary since the bandwidth depends
on the scattering potential in this model. On the other hand, for a further
understanding of the surprising amplification effect on the periodic system, it
is interesting to study the amplification effect on the resonant tunnelling in
a simple system of a double barrier which can give us a basis for the periodic
system.
//
\noindent{\bf Acknowledgments }

\hspace{0.33in} We would like to acknowledge the hospitality of the I.C.T.P. during
the progress of this work. H.B. also acknowledges the support of the physics department
at K.F.U.P.M..

\newpage

\begin{center}
{\bf Figure Captions}
\end{center}

\bigskip

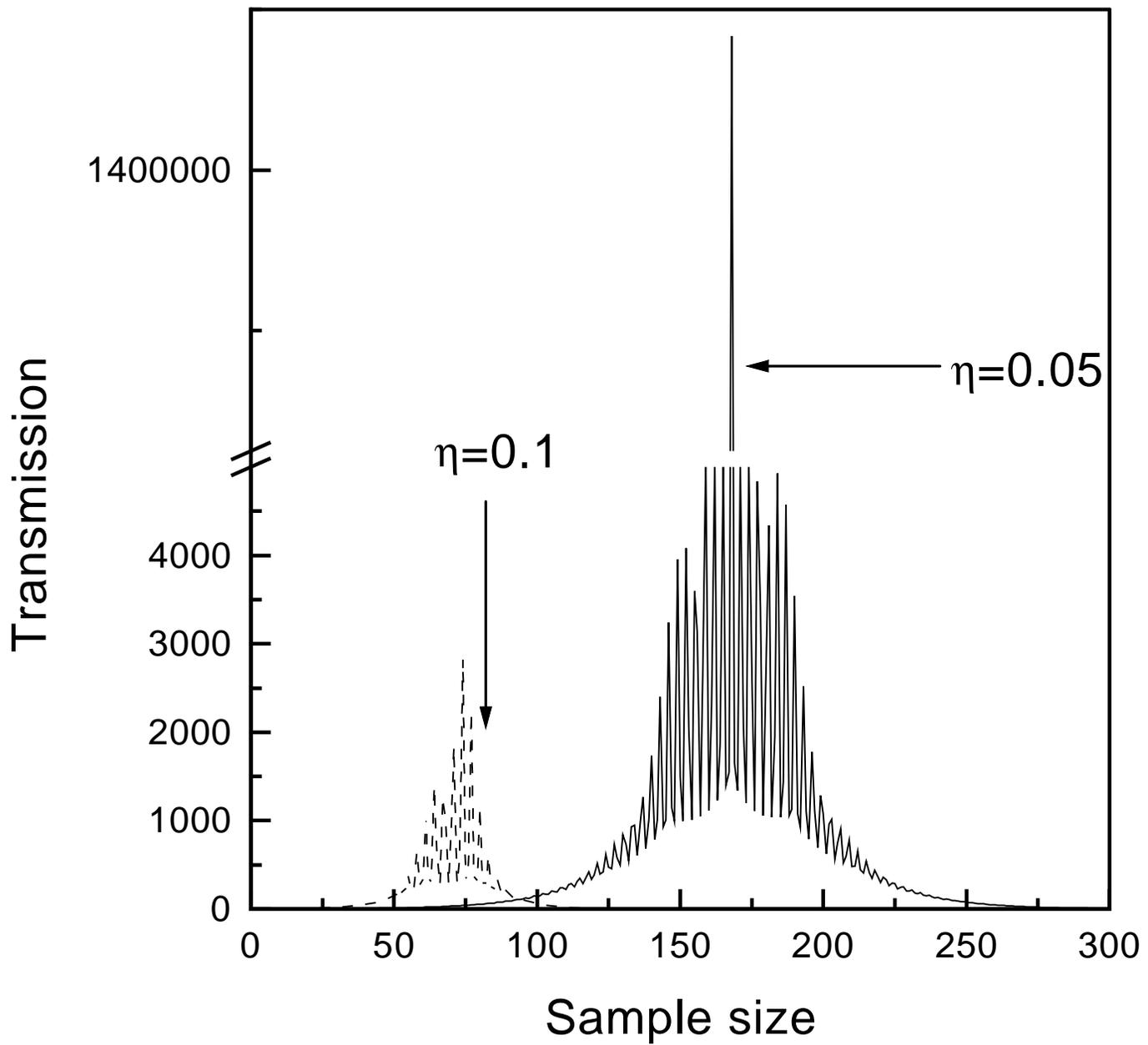
{\bf Fig.1 } Transmission coefficient versus the sample size $L$ for
$\eta =  0.05$ (solid curve) and $0.1$ (dashed curve).

\bigskip

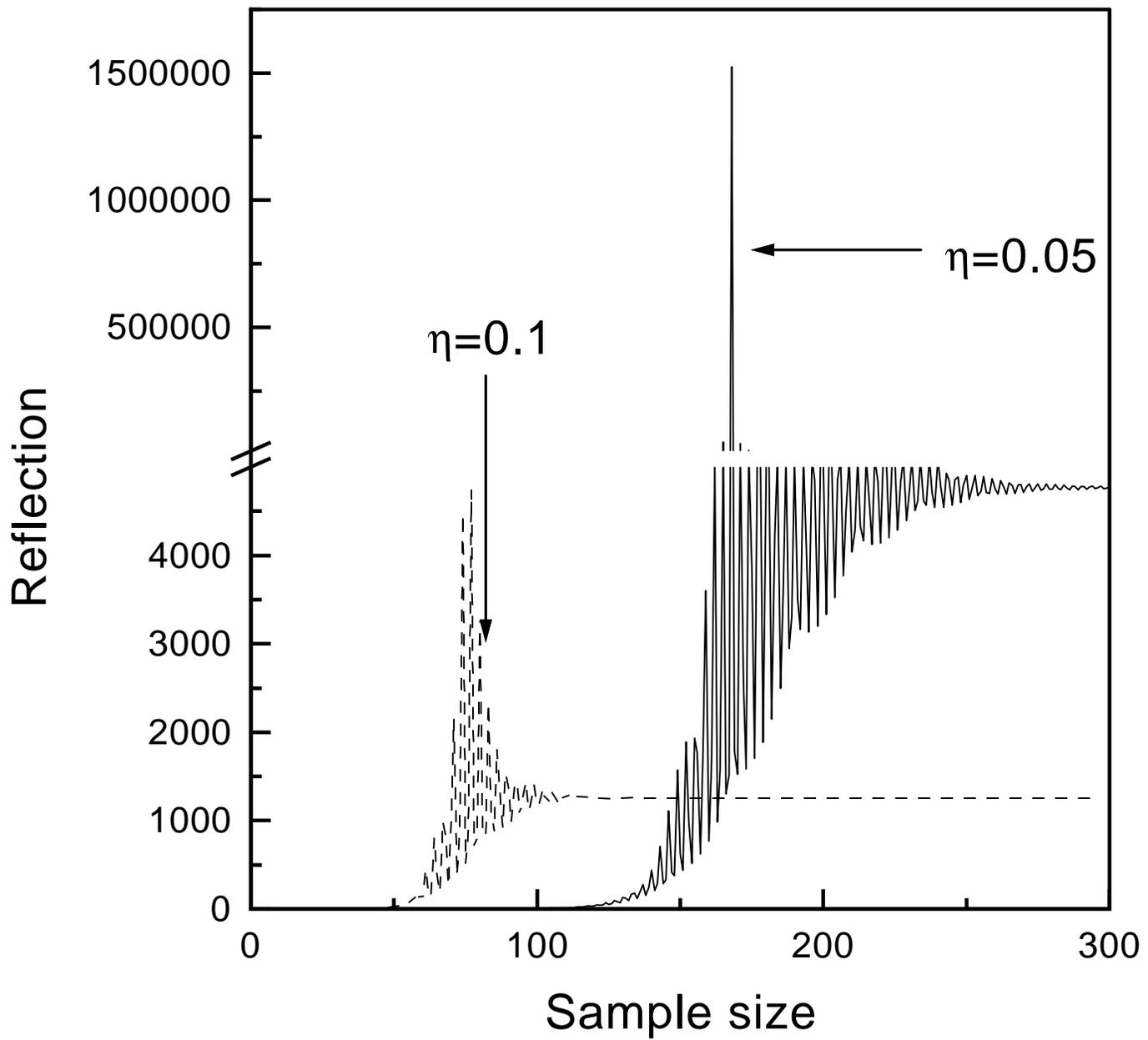
{\bf Fig.2}  Reflection coefficient versus the sample size $L$ for the
same parameters as in Fig.1 .

\bigskip

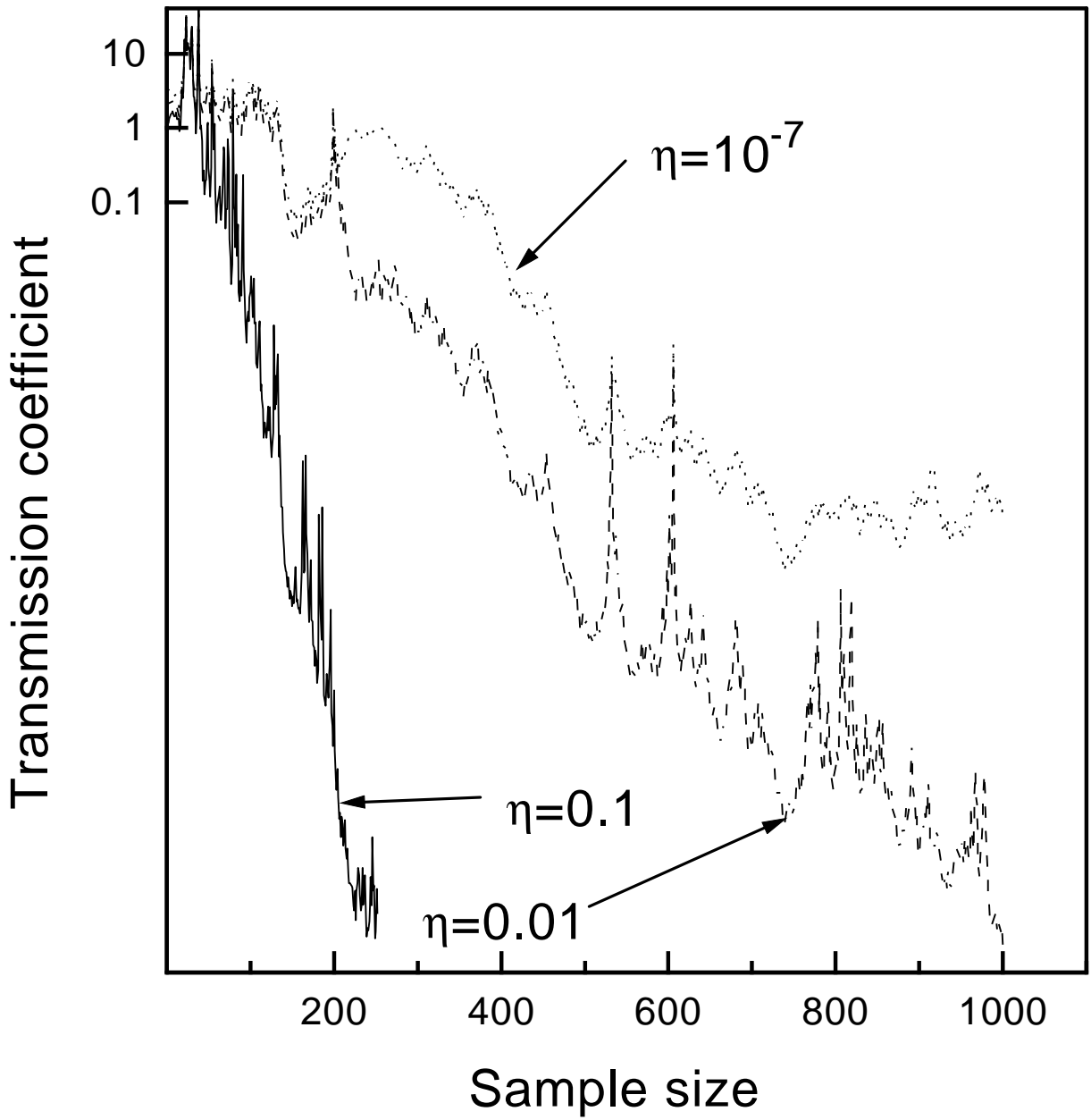
{\bf Fig.3}  $T$ versus $L$ for a disordered lattice of $\lambda_{0}$
uniformly distributed between -1/2 and 1/2 ($W=1$) and $\eta =  0.1$ (solid curve)
$0.01$ (dashed curve) and $10^{-7}$ (dotted curve).

\bigskip

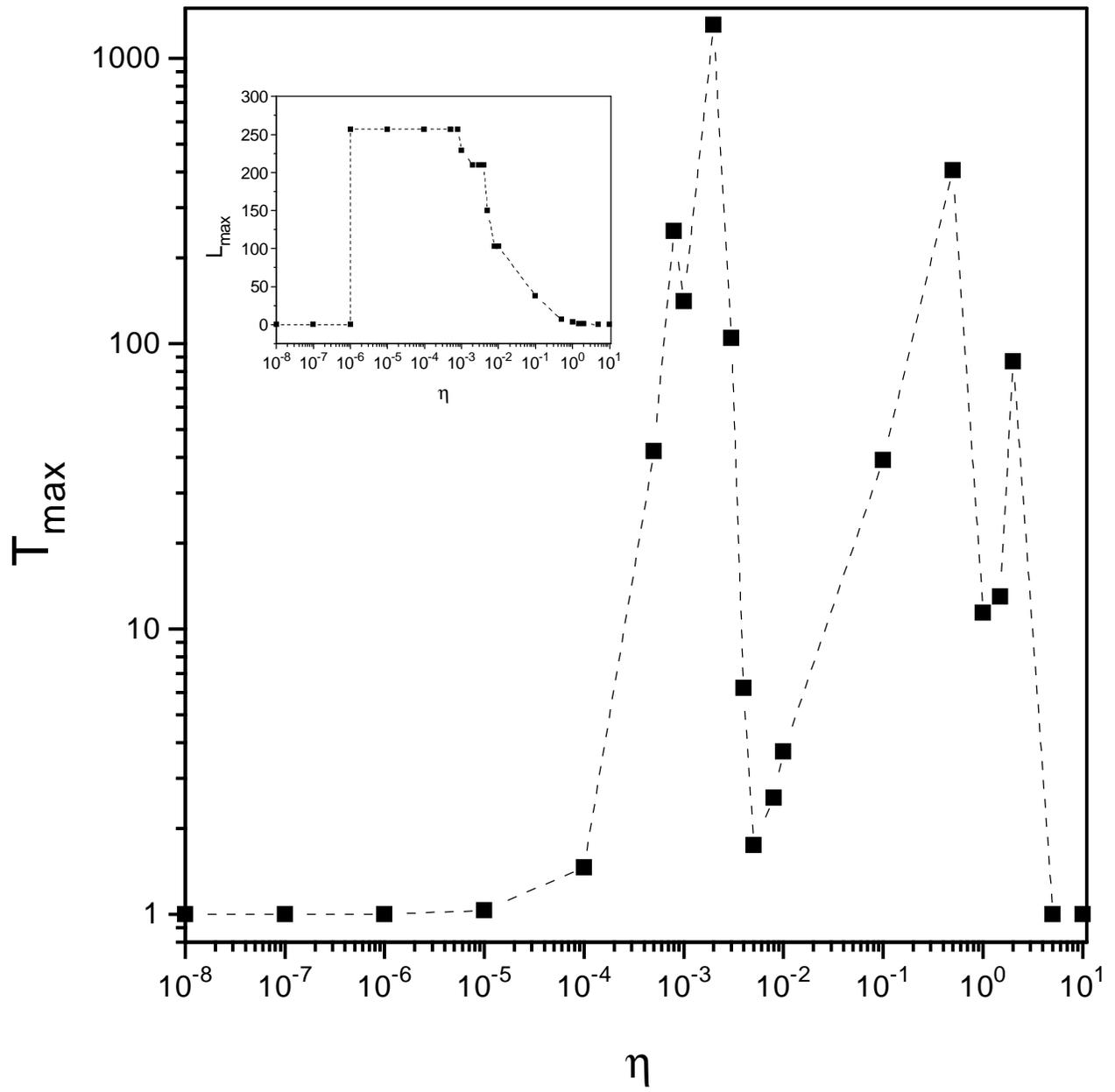
{\bf Fig.4}  $T_{max}$  versus $(\eta)$ for the same configuration of the
random real potential as in Fig.3 . The insert shows the corresponding length
at the maximum transmission ($L_{max}$) as a function of $(\eta)$. The dashed
curve is only a guide for the eyes.

\bigskip

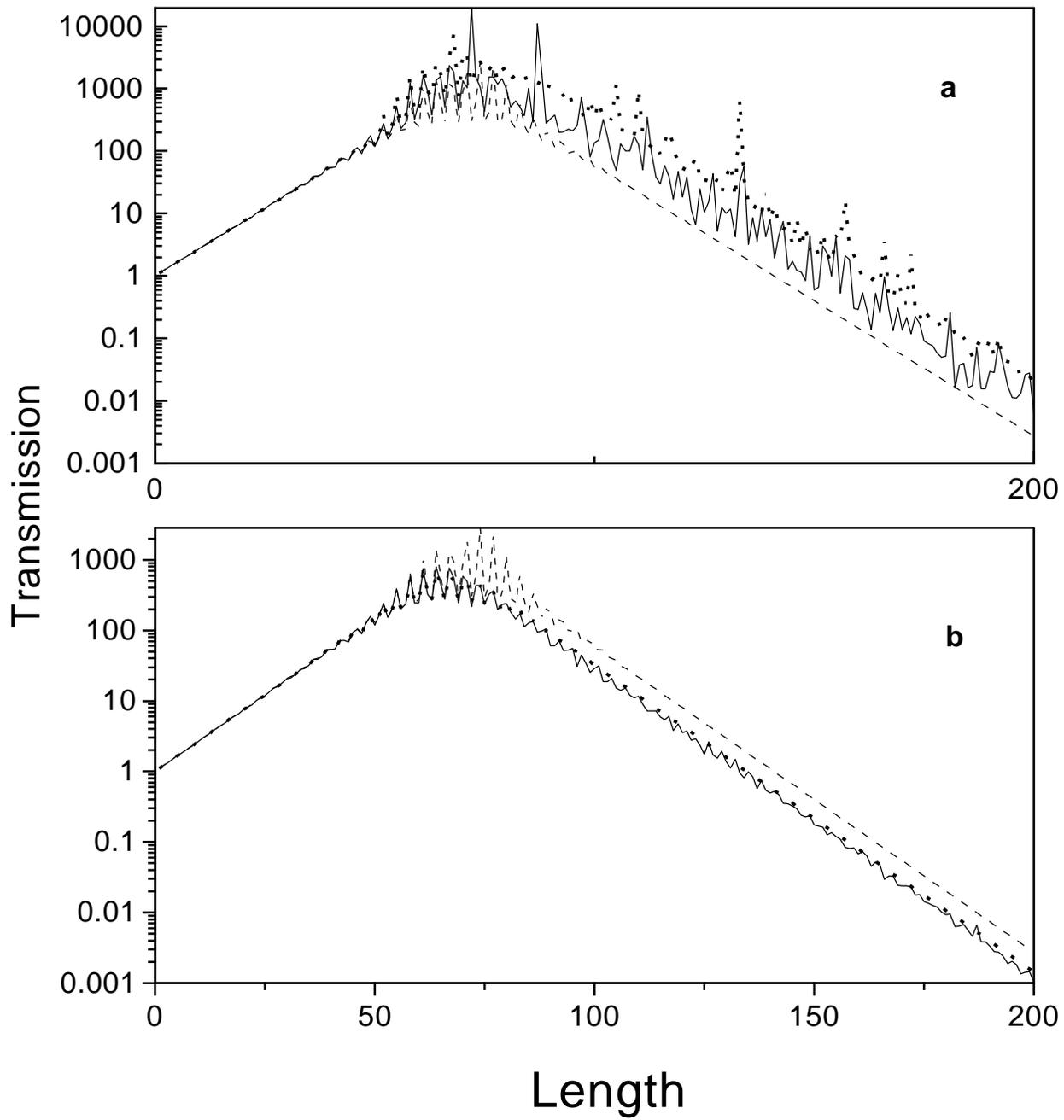
{\bf fig.5} Transmission versus length for $E=1$, $W=0.01$ and
$\eta=0.1$ for an averaging over 100 samples (solid curve), over 10000 samples 
(dotted curve) and without disorder (dahsed curve). 
 (a) averaging the quantity $T$ itself, (b) averaging ln($T$).

\bigskip

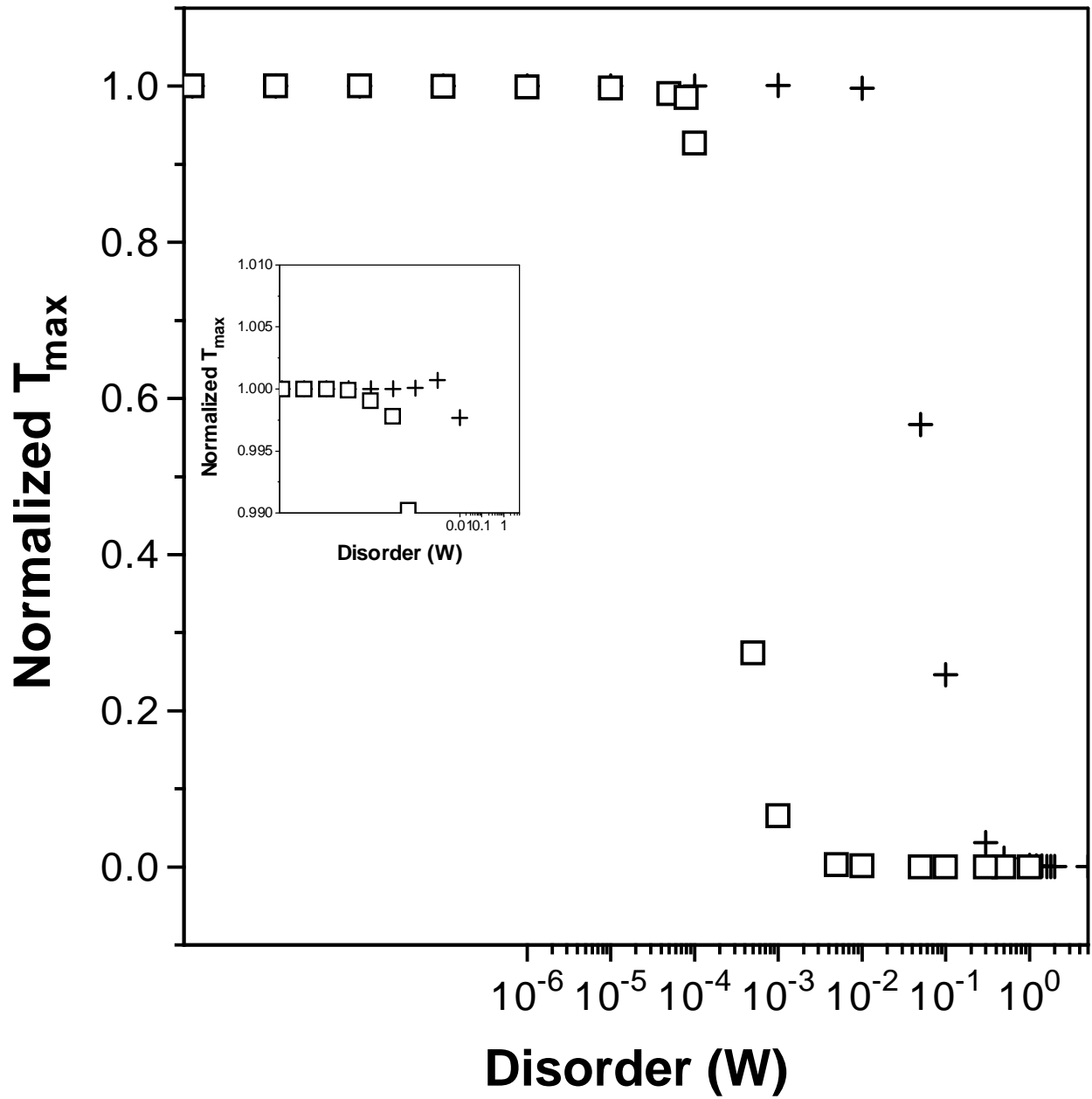
{\bf fig.6} The normalized-averaged maximum Transmission $T_{max}$ versus disorder
for $\eta=0.1$ (cross '+') and $\eta=0.01$ (open squares). The inset shows a blown up 
y-axis region between 0.99 and 1.01.

\end{document}